\documentclass[12pt]{article}
\usepackage{amsmath}
\usepackage{amsthm}
\usepackage{amsfonts}
\usepackage{amssymb}

\input epsf

\def \C {\mathbb{C}}

\def \a {\alpha}

\def \l {\ell}

\def \q  {\lq\lq}

\begin{document}

\title{Topological Quantum Computation}
\author{Michael H. Freedman$^{\ast}$, Alexei Kitaev$^{\ast\dag}$,
Michael J. Larsen$^{\ddag}$ \\ and
Zhenghan Wang$^{\ddag}$}
\maketitle $\ast$ {\it Microsoft Research, One Microsoft Way, Redmond, WA 98052}\\
$\ast\dag${\it On leave from L.D. Landau Institute for Theoretical Physics,
Kosygina St. 2, Moscow, 117940, Russia
} \\
$\ddag$ {\it Indiana University, Department of Mathematics,
Bloomington, IN 47405}

\begin{abstract}
The theory of quantum computation can be constructed from the
abstract study of anyonic systems.  In mathematical terms, these
are unitary topological modular functors. They underlie the Jones
polynomial and arise in Witten-Chern-Simons theory.  The braiding
and fusion of anyonic excitations in quantum Hall electron liquids
and 2D-magnets are modeled by modular functors, opening a new
possibility for the realization of quantum computers. The chief
advantage of anyonic computation would be physical error
correction:  An error rate scaling like $e^{-\a\l}$, where $\l$ is
a length scale, and $\alpha$ is some positive constant. In
contrast, the $\q$presumptive" qubit-model of quantum computation,
which repairs errors combinatorically, requires a fantastically
low initial error rate (about $10^{-4}$) before computation can be
stabilized.
\end{abstract}

Quantum computation is a catch-all for several models of
computation based on a theoretical ability to manufacture,
manipulate and measure quantum states.  In this context, there are
three areas where remarkable algorithms have been found: searching
a data base $\it(15)$, abelian groups (factoring and discrete
logarithm) $\it(19, 27)$, and simulating physical systems $\it(5, 21)$.
To this list we may add a fourth class of algorithms which yield
approximate, but rapid, evaluations of many quantum invariants of
three dimensional manifolds, e.g., the absolute value of the Jones
polynomial of a link $L$ at certain roots of unity:
$|V_{L}(e^{\frac{2\pi i}{5}})|$.  This seeming curiosity is
actually the tip of an iceberg which links quantum computation
both to low dimensional topology and the theory of anyons; the
motion of anyons in a two dimensional system defines a braid in
$2+1$ dimension. This iceberg is a model of quantum computation
based on topological, rather than local, degrees of freedom.

The class of functions, {\bf  BQP} (functions computable with
bounded error, given quantum resources, in polynomial time), has
been defined in three distinct but equivalent ways: via quantum
Turing machines $\it(2)$, quantum circuits $\it(3,6)$,
and modular functors $\it(7,8)$.  The last is the subject of this
article. We may now propose a $\q$thesis" in the spirit of Alonzo
Church: all $\q$reasonable" computational models which add the
resources of quantum mechanics (or quantum field theory) to
classical computation yield (efficiently) inter-simulable
classes: there is one quantum theory of computation.  (But alas,
we are not so sure of our thesis at Planck scale energies.  Who
is to say that all the observables there must even be computable
functions in the sense of Turing?)

The case for quantum computation rests on three pillars:
inevitability--- Moore's law suggests we will soon be doing it
whether we want to or not, desirability---the above mentioned
algorithms, and finally feasibility---which in the past has been
argued from combinatorial fault tolerant protocols.  (It is a
quirk of human optimism that these petitions are usually pleaded
independently: e.g., $\q$feasibility" is not seen as a
precondition for $\q$inevitability".)

Focusing on feasibility, we present a model of quantum computation
in which information is stored and manipulated in $\q$topological
degrees of freedom" rather than in localized degrees. The usual
candidates for storing information are either localized in space
(e.g., an electron or a nuclear spin) or localized in momentum
(e.g., photon polarization.) Almost by definition (see $\q$code
subspace" below) a topological degree of freedom is protected from
local errors. In the presence of perturbation (of the system
Hamiltonian) this protection will not be perfect,  but physical
arguments suggest undesirable tunneling amplitudes between
orthogonal ground states will scale like $e^{-\alpha l}$, where
$l$ is a length scale for the system, e.g.,  the minimum
separation maintained between point-like anyonic excitations.  We
will return to this crucial point.

But let us take a step backward and discuss the standard quantum
circuit model and the presumptive path toward its physical
implementation.  To specify a quantum circuit $\Gamma$, we begin
with a tensor product ${\mathbb C}_{1}^{2}\otimes \cdots \otimes
{\mathbb C}_{n}^{2}$ of $n$ copies of $\C^2$, called {\it qubits}.
Physically, this models a system of $n$ non-interacting
spin=$\frac{1}{2}$ particles. The circuit then consists of a
sequence of $K$ $\q$gates" $U_{k}, 1\leq k\leq K, $ applied to
individual or paired tensor factors. A gate is some coherent
interaction; mathematically it is a unitary transformation on
either $\C_i^{2}$ or $\C_{i}^{2}\otimes \C_{j}^{2}$, $1\leq
i,j\leq n$, the identity on all remaining factors.   The gates are
taken from a fixed finite library of unitary $2\times 2$ and
$4\times 4$ matrices (with respect to a fixed basis $\{|0\rangle, |1\rangle\}$
for each $\C^{2}$ factor) and must obey the surprisingly mild
condition, called $\q$universality", that the set of possible gate
applications generates the unitary group ${\mathbb U}(2^{n})$
densely (up to a physically irrelevant overall phase.)  Popular
choices include a relative phase gate $ \left(
\begin{array}{cc}
1& 0\\
0 & e^{\frac{2\pi i}{5}}
\end{array}
\right) $ and $\q$Controlled NOT" $ \left(
\begin{array}{cccc}
1& 0&0& 0\\
0&1&0&0 \\
0&0&0& 1\\
0 & 0& 1& 0
\end{array}
\right) $ operating on one and two $\q$particles", respectively.
It is known that beyond the density requirement the particular
choice of gates is not too important.  Let
$W_{\Gamma}=\prod_{i=1}^{m}U_{i}$ denote the operator effected by
the circuit $\Gamma$.  It is important for the fault tolerance
theory that many gates can be applied simultaneously (to
different qubits) without affecting the output of the circuit
$W_{\Gamma}(|0\rangle\otimes \cdots \otimes |0\rangle)$.

Formally, information is extracted from the output by measuring
the first qubit. The probability of observing $|1\rangle$ is given
according to the usual axioms of quantum mechanics as:
\begin{equation}
p(\Gamma)=\langle 0|W_{\Gamma}^{\dagger}\Pi_{1}W_{\Gamma}|0\rangle,\label{(1)}
\end{equation}
where $\Pi_{1} $  is the projection to $|1\rangle$, $\left(
\begin{array}{cc}
0& 0\\ 0 & 1
\end{array}
\right) , $ applied to the first qubit.  Any decision problem,
such as finding the $k$-th binary digit of the largest prime
factor of an integer $x$, can be modeled by a function $F$ on
binary strings, $F: \{0,1\}^{*}\rightarrow \{0,1\};$ in our
example, the input string would encode $(x,k)$. We say $F$
belongs to {\bf BQP} if there is a classical polynomial-time (in
string length) algorithm for specifying a $\q$quantum circuit"
$\Gamma(y)$ (in the example $y=(x,k)$) which satisfies: $$
\begin{array}{cccc}
p(\Gamma(y))\geq \frac{2}{3} & \text{if} & F(y)=1 & \text{and} \\
p(\Gamma(y))\leq \frac{1}{3} & \text{if} & F(y)=0. &
\end{array}
$$

This definition suggests that one needs to make an individual
quantum circuit to solve each instance of a computational problem.
However, it is possible to construct a single circuit to solve any
instance of a given {\bf BQP} problem of bounded size, e.g.,
factor integers with $\leq 1000$ digits.  Moreover, by $\it(19)$ there
is a universal circuit, univ.(n,k),  which simulates all circuits
of size $k$ on $n$ qubits:
$$ W_{\text{univ.(n,k)}}\Bigl(|0 \cdots 0\rangle \otimes |\Gamma\rangle\Bigr)
=W_{\Gamma}|0\cdots 0\rangle\otimes |\Gamma\rangle.$$

Yet another definition allows one to do measurements in the middle
of computation and choose the next
 unitary gate depending on the measurement outcome.  This choice will
 generally involve some classical Boolean gates.  This scheme is called
 {\it adaptive quantum computation}.  In certain cases, it can
 squeeze general {\bf BQP} computation out of a gate set which is
 {\it not} universal.

 {\section*{Implementation of a quantum computer}}

 It is not possible to realize a unitary gate precisely, and even
 when we do not intend to apply a gate, the environment will interact
 with the qubits causing decoherence.  Both imprecision and decoherence
 can be considered in a unified way as $\q$errors" which can be quantified
 by a fidelity distance $\it(17)$ or a super-operator norm $\it(19)$.
 A crucial step in the
 theory of quantum computing has been the discovery of error-correcting
 quantum codes $\it(28)$ and fault-tolerant quantum computation $\it(25, 29)$.
 These techniques cope with sufficiently small errors.  However,  the
error
  magnitude must be smaller than some constant (called {\it an accuracy
  threshold}) for these methods to work.  According to rather optimistic
  estimates, this constant lies between $10^{-5}$ and $10^{-3}$,
  beyond the reach of current technologies.

  The presumptive path toward a quantum computer includes
  these steps: (1) build physical qubits and physical gates; (2) minimize
  error level down below the threshold; (3) implement
decoherence-protected
  logical qubits and fault-tolerant logical gates (one logical qubit is
  realized by several qubits using an error-correcting code.)  As a
counterpoint,
  the theme of this article is that implementing physical qubits might be
  redundant.  Indeed, one can $\q$encode" a logical qubit into a physical
system
  which is not quite a set of qubits or even well separated into
subsystems. Such a system must have macroscopic quantum degrees of
freedom which would
  be decoherence-protected.  A super-conducting phase or anything related
to
  a local order parameter will not work for if a degree of
freedom is accessible by local measurement, it is vulnerable to
local error.  However, there is another
  type of macroscopic quantum degree of freedom.  It is related to
topology
  and arises in collective electronic systems, e.g. the fractional quantum
  Hall effect $\it(13)$
  and most recently in $2D$ cuprate superconductors above $T_c$ $\it(12, 23)$.

   Though much
  studied since the mid-1980's the connection between fractional
  quantum Hall effect and quantum
computation
  has only recently been realized $\it(11, 20)$.
  It was shown by $\it(4)$ that the ground state
  of the $\nu=\frac{1}{3}$ electron liquid on the torus is 3-fold
degenerate.
  This follows from the fact that excitations in this system are abelian
anyons:
 moving one excitation around another multiplies the state vector
 by
 a phase factor $e^{i\phi}$ (in this case $\phi=\frac{2\pi}{3}$).
The process of creating
  a particle-antiparticle pair, moving one of the particles around the
  torus,
  and annihilating it specifies a unitary operator on the ground state.
By
  moving
the
  particle in two different directions, one obtains two different unitary
operators
  $A_1$ and $A_2$ with the commutation relation
  $A_1A_2A_{1}^{-1}A_{2}^{-1}=e^{i\phi}$, implying a ground state
degeneracy. This
  argument is very robust and only requires the existence of an energy gap
  or, equivalently, finite correlation length $l_0$.  Indeed, the
degeneracy is
  lifted only by spontaneous tunneling of virtual excitations around the
torus.
  The resulting energy splitting scales as $e^{-\frac{l}{l_0}}$, where
  $l$ is the size of the system.  Interaction with the environment does
not
  change this conclusion, although thermal noise can create actual
excitation
  pairs rather than virtual ones.

  Both the ground state degeneracy on the torus and the existence of
anyons are
  manifestations of somewhat mysterious topological properties of the
  $\nu=\frac{1}{3}$ electron liquid itself.  Anyons can be regarded as
  $\q$topological defects" similar to Abrikosov vortices but without any
local
  order parameter.  The presence of a particle
enclosed by a loop
  on the plane can be detected by holonomy---moving another particle
around the loop.

  At $\nu=\frac{1}{3}$, the electron liquid on the torus could be used as
a logical
  $\q$qutrit" (generalized qubit with 3 states).  Unfortunately, this will
hardly
  work in practice.  Besides the obvious problem with implementing the
torus
  topology,  there is no known way to measure this logical qutrit or
prepare
  it in a pure state.

  A more flexible and controllable way of storing quantum information is
based
  on nonabelian anyons.  These are believed to exist in the
  $\nu=\frac{5}{2}$ fractional quantum Hall state.
  According to the theory $\it(22, 24)$,  there
should
  be charge $\frac{1}{4}$ anyonic particles and some other excitations.
The
  quantum state of the system with $2n$ charge $\frac{1}{4}$ particles on
the plane
  is $2^{n-1}$-degenerate.  The degeneracy is gradually lifted as two
  particles come close to each other.  More precisely, the
  $2^{n-1}$-dimensional Hilbert space $\mathbb H_n$ splits into two
  $2^{n-2}$-dimensional
  subspaces.  They correspond to two different types of charge
$\frac{1}{2}$
  particles which can result from fusion of charge $\frac{1}{4}$
particles.
  Thus observing the fusion outcome effects a measurement on the Hilbert
space
  $\mathbb H_n$.  This model supports adaptive quantum computation when
surfaces of high
  genus are included in the theory and admits a combinatorial description $\it(1)$
  apparently in the same universal class as the fractional quantum Hall
fluid.

  Beyond this, a discrete family of quantum Hall models exists $\it(26)$ based
on
  $k+1$-fold hard-core interaction between electrons in a fixed Laudau
level which
  appears to represent the same universality class as
   Witten-Chern-Simons theory for $SU(2)$ at level=$k$  $\it(33)$.
  Anyons in these models behave as topological defects
  of a geometric construction $\it(10)$ and their braiding matrices
   have been shown to be universal $\it(8,9)$
  for $
  k\geq 3, k\neq 4$.

   {\section*{Code spaces and quantum media}}

  Even after the particle types and positions of anyons are
  specified, there is an exponentially large (but finite
  dimensional) Hilbert space describing topological degrees of
  freedom.  In combinatorial models, this Hilbert space becomes a
  $\q$code subspace" $W$ of a larger $\q$quantum media" $Y$.  Thus
  a fundamental concept of cryptography is transplanted into
  physics.  Let $V$ be a finite dimensional
  complex vector space, perhaps $\C^{2}$, and $Y=V\otimes \cdots \otimes
V$
  an $n$-fold tensor product (where $n$ is typically quite large.)  Let
$W\subset Y$
  be a linear subspace.  We call $W\subset Y$ $k$-code if and only if:
  $$W\stackrel{\Pi_{W}\cdot \mathcal O}\longrightarrow W$$
  is multiplication by a scalar whenever $\mathcal O$ is a $k$-local
operator (an
  arbitrary linear map on any $k$-tensor factors of $Y$ and the identity
  on the remaining $n-k$ factors ) and $\Pi_{W}$ is the orthogonal
projection
  onto $W$.  We think of such spaces as resisting local alteration and
  in the usual interpretation of $Y$ as the Hilbert space of $n$
particles,
  quantum information stored in $W$ will be relatively secure.  It
  is a theorem $\it(14)$ that the quantum information in $W$ cannot be degraded
by errors
  operating on fewer than $\frac{k}{2}$ of the $n$ particles.

  Let us define a (discrete) quantum medium to be a tensor product
  $Y=\otimes_{i}V_{i}$ as above, where now the set of indices  $\{i\}$
  consists of points distributed on a geometric surface $T$, together with
  a local Hamiltonian $H=\sum H_i$ (each $H_i$ is a Hermitian operator
  defined only on those tensor factors whose index is within
  $\epsilon \rangle0$ of the $i$-th point in the geometry of the surface.)  Local
Hamiltonians  $H$
  have been found $\it(10,20)$ with highly $d$-degenerate ground
  states
  corresponding to modular functors $\it(31,32)$, (and thus
  braid group representation $\it(16)$ and link polynomials $\it(18)$.)
  In these cases, the ground state $G$ of $H$ will be $k$-code for
  $k\sim \text{injectivity radius of
T}\sim \sqrt{\text{area T}}$.  The topological degrees of freedom
referred to above reside in $G$.  But we do not attempt a precise
mathematical definition of topological degrees of freedom  since
we would like it to extend beyond discrete system, e.g., to
fractional quantum Hall ground states.

  In the case
  when $T$ is a disk $D$ with punctures---physically {\it anyonic
excitations}---a
   sequence
  of local modifications to $H$ (see $\it(10)$) effects a discrete 1-parameter
family
  $H_t$ of Hamiltonians, where the ground states $G_{t_{i}}$ and
$G_{t_{i+1}}$ at
  consecutive time steps differ by a $\lfloor \frac{k}{2} \rfloor$-local
operator
  $\mathcal O_i$, $\mathcal O_i(G_{t_{i}})=G_{t_{i+1}}$.
  Note that if $G_{t_{i}}$ and $G_{t_{i+1}}$
  are both $k$-code that for $\mathcal O_i$ as above,
  $\mathcal O_i|_{G_{t_{i}}}$ must be unique up to a
scalar
  (for a distinct $\mathcal O_i'$, consider
  the restriction of the $k$-local
  operator ${\mathcal O_i}^{\dag}\circ \mathcal O_i^{'}$
  projected to $G_{t_i}$, $\Pi_{G_{t_{i}}}\circ {\mathcal
  O_i}^{\dag}\circ
  \mathcal O_i^{'}$).  This uniqueness property forces this
  discrete-cryptographic transport of code spaces to coincide (up to
phase) with
  the differential geometric notion of adiabatic transport---integration
  of the canonical connection in the $\q$tautological" bundle of
  $d$-planes in $Y=\C^{2^{n}}$.  If the 1-parameter family is a closed
loop, a
  projective representation of the braid group on $\mathbb U(\text{code})$
  is obtained.
  By choosing $H$, these can be engineered to be
  precisely the Hecke algebra representations $\{\rho_{\lambda}\}$
  associated to the Jones polynomial.  From $H$ one
  can build a concrete model of quantum computation; the model
  and its connection to the Jones polynomial are described below.
Although
  the $H$ found in $\it(10)$ is enormously cumbersome, it appears to share the
  universality class $\q$Witten-Chern-Simons theory of
  $SU(2)$ at level $3$" with a simple $4$-body Hamiltonian $\it(26)$, which has
been
  proposed as a model for the fractional quantum
  Hall plateaus at $\nu=\frac{13}{5}\;\text{and}\; \frac{12}{5}$.

  {\section*{An anyonic  model for quantum computation}}

  A family of unitary representations of all mapping class and braid
  groups with certain compatibility properties under fusion
  is known as a unitary topological modular functor.
  To define our model, we take
only
  the planar surface portion of the simplest universal modular functor,
  Witten-Chern-Simons $SU(2)$ modular functor at level 3,
  and this reduces to the Jones
representations
  of braids $\{\rho_{\lambda} \}$.
   For an appropriate family of local Hamiltonians $H_t$, these
representations describe
  the adiabatic transport of the lowest energy states with $n$ anyonic
  excitation pairs, these states form a subspace $W$ of dimension
Fibonacci(2n)---
  as the $2n$ anyonic excitations are $\q$braided" around each other in
2+1 dimensional
  space time.  In this theory, (we denote it $CS5$ because of its link
  to the Jones polynomial at a fifth root of unity), there are 4 label
types
  0,1,2,3 corresponding to the complete list of irreducible
  representations of the quantum group $SU(2)_{5}$ of dimensions 1,2,3 and
4 (or equivalently
  the irreducible positive energy loop group representations
  of $\mathcal L SU(2)$ at level $3$).
  We initialize our system on the disk $D$ in a
  known state by pulling anyonic pairs out of the vacuum.  This theory is
self-dual so
  the
  two partners have identical types.  Pairs are kept or returned to the
vacuum according
  to the results of local holonomic measurement.  Finally we have a known
  initial state in the disk with $2n$ punctures where each puncture has
  label=1 and $\partial D$ has label=0.  We assume $n$ is even
  and group the punctures into $n/2$ batches of 4 punctures each.
  Similar to $\it(8)$, each batch $B$ is used to encode one qubit
  $\cong \C^{2}$: the basis $\{|0\rangle, |1\rangle\}$ is mapped into the
  type ($0$ or $2$) of the 2-fold composite particle
   (round circle),  which would result from fusing a
(fixed)
  pair of the type 1 particles within $B$.  Initially,  both the
  double and 4-fold composites (ovals) have type 0.  By
  maintaining, at least approximately, this condition on the ovals
  after the braiding is complete, we define a $\q$computational summand"
   of the modular functor:  it is spanned by $n$-bit strings of
   $0$'s and $2$'s residing on the 2-fold composites (round
   circles).

  \vskip.1in \epsfxsize=3in \centerline{\epsfbox{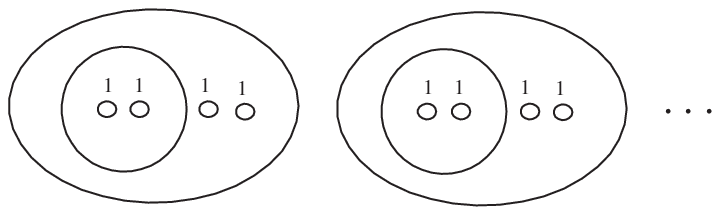}}
\centerline{Figure 1} \vskip.2in

 Now as in the quantum circuit model, a classical poly-time algorithm
  looks at the problem instance $(F,y)$ and builds a sequence of
$\q$gates", but
  this time the gates are braid generators
   (right half twist between adjacent anyons)
   $\sigma_i \; ,\; 1\leq i\leq 2n-1$,  and a powerful approximation
  theorem $\it(19,30)$ is used to select the braid sequence which
approximates the more
  traditional quantum circuit solving $(F,y)$.  So the topological model
  may be described as:

\begin{enumerate}
  \item Initialization of a known state in the modular functor.

  \item Classical computation of braid $b$ effecting a desired unitary
  transformation $X$ of the computational subspace of the modular
  fucntor.

  \item Adiabatic implementation of the braid by (somehow) moving the
anyons
  in $D$ to draw $b$ in space-time. (Here we keep
  the anyons separated by a scale $l$.)

   \item Application of  a projection operator
  $\Pi$ to measure the type ($0$ or $2$)
  of the $\q$left most" composite particle (as seen in Figure 1.).
\end{enumerate}

  The last step is the direct analogue of measuring the first qubit
  in the quantum circuit model and the same formula (1) applies:
  the probability of
  observing type 0 (the null particle) is:
 $$\text{prob}(0)=\langle 0|X^{\dagger}\Pi X|0\rangle,$$
  where the $0$'s on the right hand side represent our carefully prepared
  initial state with $2n$ type 1 excitations.

  To close the topological discussion, we note that the previous formula
   can be translated (using the $S$ matrix) into a plat closure (See Figure 2) of
   a braid $L=\text{plat}(b^{-1}\gamma b)$, where $\gamma$ is a small
   loop inserted to measure the left-most qubit, and now the outcome of
the
   quantum circuit calculation, $\text{prob}(0)$, becomes a Jones
   evaluation

   \vskip.2in \epsfxsize=3in \centerline{\epsfbox{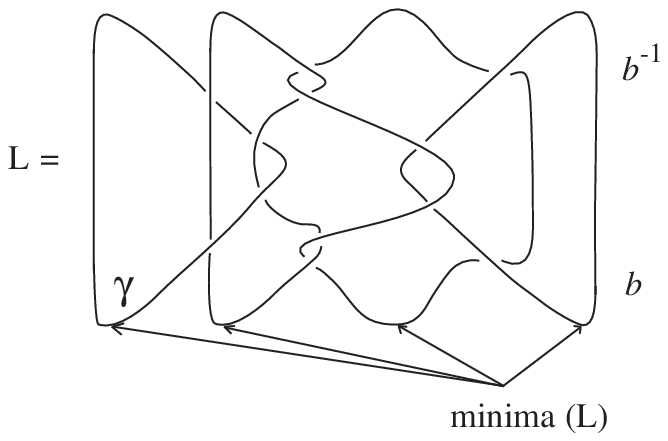}}
\centerline{Figure 2} \vskip.2in

  \[
   \text{prob} (0)=\frac{1}{1 + [2]^2_5}
   \left(1+\frac{(-1)^{c(L) + w(L)} (-a)^{3 w(L)} V_L(e^{2 \pi i/5})}{[2]_5^{m(L)-2}}\right);
  \]
where $[2]_5 =\frac{1+\sqrt{5}}{2}$; $c, m$, and $w$ are the
number of components, number of local minima, and writhe of $L$
respectively; and $a= e^{\pi i/10}$. Details of this calculation
will be posted at url: www.tqc.iu.edu.

  The braiding disturbs and reforms composite particle types with
sufficient
  subtlety to effect universal computation.
 To reduce this model to engineering, very significant obstacles must
   be overcome: stable quantum media must be maintained in
   a suitable phase, e.g.,
   $CS5$; excitations must be readily manipulated, and electrical neutral
   particle types 0 and 2 must be distinguished,
   presumably by holonomy experiments.  Although these
   challenges are daunting, they are, perhaps, less difficult than a
head-on
   assault on the accuracy threshold in the quantum circuit model.

\end{document}